\documentclass[prd,nofootinbib,preprintnumbers,preprint,
floatfix]{revtex4}
\usepackage{empheq}

\usepackage{amsmath}
\usepackage{amsfonts}
\usepackage{amssymb}
\usepackage{dsfont}
\usepackage{bm}
\usepackage[hyperfootnotes=false]{hyperref}
\usepackage[dvipsnames]{xcolor}
\usepackage{apptools}
\usepackage{enumerate}
\usepackage{subcaption}
\usepackage{diagbox}

\begin{document}

\title{Perturbation theory for gravitational shadows in Kerr-like spacetimes}
\author{Kirill Kobialko${}^{1,\,}$\email{kobyalkokv@yandex.ru}}
\author{Dmitri Gal'tsov${}^{1,\,}$\email{galtsov@phys.msu.ru}}
\affiliation{${}^1$ Faculty of Physics, Moscow State University, 119899, Moscow, Russia}

\begin{abstract}
We present a fully analytical method for calculating the key parameters of a Kerr-like gravitational shadow, including its horizontal and vertical diameters, $D_X$ and $D_Y$, the coordinates of its center $X_{C}$, the average radius $\bar{R}$, the deviation from sphericity $\delta C$, and the mean deviation from the Kerr shadow $\delta K$. Developed within the framework of perturbation theory, this approach yields all characteristic parameters as simple polynomial expressions with an accuracy of $\sim a^5$, where $a$ is the Kerr spin parameter. This eliminates the need for repeated numerical integration of cumbersome parametric equations. Furthermore, our derived formulas account for the effects of a plasma medium—a feature of particular relevance given the prospect of multi-frequency astrophysical observations.
\end{abstract}

\maketitle

\setcounter{page}{2}

\setcounter{equation}{0}
\setcounter{subsection}{0}

\section{Introduction} 

Recent observations of black hole shadows by the Event Horizon Telescope collaboration \cite{EventHorizonTelescope:2019dse,EventHorizonTelescope:2022wkp,Akiyama:2024zpm} have provided significant impetus for the development of novel theoretical approaches to calculating and analyzing gravitational shadows. In these studies, photon spheres and their generalizations play a central role, as they define the boundaries separating particle trajectories that escape to infinity from those captured by the black hole \cite{Virbhadra:1999nm,Virbhadra:2002ju,Virbhadra:2008ws,Shoom:2017ril,Gibbons:2016isj,Perlick:2021aok,Cunha:2018acu,Grenzebach:2014fha,Grenzebach:2015oea,Stepanian:2021vvk}. These structures are most straightforwardly defined for photons in static spacetimes \cite{Claudel:2000yi}. In rotating geometries, an analogous role is played by surfaces that host non-planar spherical orbits \cite{Teo:2020sey,Kobialko:2020vqf}. The concept naturally extends to massive particles and to particles with variable effective mass, such as photons in a dispersive plasma \cite{Perlick:2015vta,Perlick:2017fio,Perlick:2023znh,Bezdekova:2022gib,Briozzo:2022mgg,Bogush:2023ojz,Kobialko:2022uzj,Song:2022fdg,Bogush:2024fqj}. As in the vacuum photon case, this framework enables the systematic derivation of analytical expressions for gravitational shadows \cite{Kobialko:2023qzo}, whose morphology inherently depends on the particle's energy (or observational frequency). This energy-dependent nature of the shadow allows for the extraction of additional information from observational data \cite{Moscibrodzka:2017lcu,Chael:2022meh,Ricarte:2022sxg}, thereby establishing the foundation for the shadow spectroscopy \cite{Pantig:2025deu}. 

However, even with an analytical expression available \cite{Kobialko:2023qzo}, a comprehensive analysis of the gravitational shadow remains a challenging task. This difficulty arises because the shadow boundary is typically parameterized by the radius of the spherical photon orbit, rather than directly by the angle on the observer's celestial sphere. As a result, calculating the standard shadow parameters \cite{Johannsen:2013vgc,Cunha:2015yba} — such as the horizontal and vertical diameters $D_X$ and $D_Y$, the center coordinate $X_{C}$, the average radius $\bar{R}$, the deviation from sphericity $\delta C$, and the mean deviation from the Kerr shadow $\delta K$—in a fully analytical manner is non-trivial. In practice, these characteristics usually require numerical computation even when an exact parametric form of the shadow boundary is known. Since performing such numerical calculations over wide ranges of metric parameters can be cumbersome, having reliable analytical approximations is particularly valuable. A series of recent studies, primarily on spherically symmetric spacetimes, has demonstrated the efficacy of a perturbation-theoretic approach \cite{Vertogradov:2024dpa,Vertogradov:2024eim,Pantig:2025deu,Pantig:2025eqe,Kobialko:2024zhc,Mishra:2019trb}. These works have also underscored the necessity for generalizing such methods to stationary, Kerr-like geometries \cite{Pantig:2025deu}.

We present an analytical approach for calculating key shadow parameters in Benenti-Francavilla-type \cite{Papadopoulos:2018nvd,Benenti:1979erw,Demianski:1980mgt} stationary metrics using perturbation theory. This method yields key shadow characteristics, such as the diameters and the center position, as simple polynomial expansions accurate to $\sim a^5$, where $a$ is the Kerr spin parameter, thereby bypassing the need for repeated numerical integration of cumbersome parametric equations. Benenti-Francavilla metrics represent a broad class of geometries that include many Kerr-like solutions—such as Kerr-off shell metric \cite{BenAchour:2025uzp} — and are characterized by the complete separation of variables in the geodesic equations \cite{Papadopoulos:2018nvd,Konoplya:2021slg}. Our formalism also incorporates the effects of a plasma medium, a feature of significant relevance for multi-frequency astrophysical observations. The availability of such analytical expressions serves a dual purpose: it provides direct insight into how each additional model parameter influences the shadow's morphology and significantly simplifies the inverse problem of recovering these parameters from observational data using standard optimization techniques.

The paper is structured as follows. In Section \ref{sec:setup}, we derive a general expression for the gravitational shadow boundary in integrable Benenti-Francavilla-type metrics. Section \ref{sec:perturbation} presents the derivation of the key shadow parameters using perturbation theory around the Schwarzschild metric. The application of these general formulas to several specific metrics is discussed in Section \ref{sec:examples}, demonstrating both the high accuracy of the approximation and its utility for metric parameter reconstruction. The main results are summarized in the Conclusion, and an Appendix provides supplementary expressions useful for shadow analysis.
  
\section{Setup} 
\label{sec:setup}

This section derives a general expression for the gravitational shadow boundary as perceived by a static asymptotic observer within the framework of Benenti-Francavilla-type metrics \cite{Papadopoulos:2018nvd,Papadopoulos:2020kxu,Benenti:1979erw}. While expressions of this kind are available in various forms in the literature \cite{Grenzebach:2014fha,Grenzebach:2015oea,Konoplya:2021slg,Perlick:2021aok,Banerjee:2019nnj,Vincent:2020dij,Tsukamoto:2014tja,Tsukamoto:2017fxq}, they provide the necessary foundation for our subsequent perturbation-theoretic analysis. From the outset, we consider the general case of radiation propagating through a dispersive plasma medium \cite{Perlick:2015vta,Perlick:2017fio,Perlick:2023znh,Bezdekova:2022gib}.

In the geometrical optics approximation, the propagation of radiation in a cold, non-magnetized, and pressureless plasma is governed by the Hamiltonian \cite{Perlick:2023znh,Kichenassamy:1985zz}
\begin{align} \label{eq:hm_a}
H=\frac{1}{2}(g^{\alpha\beta} p_\alpha p_\beta + \omega^2_p), 
\end{align}
where $p_\alpha$ are the photon generalized momenta, $g^{\alpha\beta}$ is the inverse spacetime metric, and $\omega_p$ is the plasma frequency. We consider a spacetime metric in the Benenti-Francavilla \cite{Papadopoulos:2018nvd,Benenti:1979erw} form and assuming a separable plasma distribution \cite{Perlick:2017fio,Kobialko:2023qzo}. In the coordinates $x^\alpha = (r, \theta, t, \phi)$, the metric and the plasma frequency distribution are given by
\begin{align} \label{eq:hm}
g^{\alpha\beta}=\Sigma^{-1}\begin{pmatrix}
A_2 & 0 & 0 & 0 \\
0 & B_2 & 0 & 0 \\
0 & 0 & A_3+B_3 & A_4+B_4  \\
0 & 0 & A_4+B_4 & A_5+B_5 
\end{pmatrix}, \quad \omega^2_p=\frac{f_r(r)+f_\theta(\theta)}{\Sigma}, \quad \Sigma=A_1+B_1,
\end{align}
where $A$ and $B$ are functions of $r$ and $\theta$, respectively. This class of metrics permits complete separation of variables in the Hamilton-Jacobi equations \cite{Papadopoulos:2018nvd,Benenti:1979erw} and is algebraically general \cite{Galtsov:2024vqo}. It therefore provides a broader parametrization of Kerr-like spacetimes than, for instance, the Kerr-off shell geometry \cite{BenAchour:2025uzp}. 

Specifically, starting from the Hamilton-Jacobi equation
\begin{align} 
g^{\alpha\beta}  \partial_\alpha S \cdot\partial_\beta S+ \omega^2_p=0,
\end{align}
we obtain two decoupled equations
\begin{align} 
B_2p_\theta^2=\mathcal{J}\equiv\mathcal{C}-B_3\omega^2+2B_4\omega L-B_5 L^2-f_\theta, \\
A_2p_r^2=\mathcal{R}\equiv-\mathcal{C}-A_3\omega^2+2A_4\omega L-A_5L^2- f_r. \label{eq:Hamilton-Jacobi}
\end{align}
where $-\omega=p_t =\partial_t S$ and $L=p_\phi = \partial_\phi S$ correspond to the conserved energy and angular momentum of the photon, $p_r = \partial_r S$ and $p_\theta = \partial_\theta S$ are its radial and $\theta$-angular momenta, and $\mathcal{C}$ denotes the Carter constant. We also recall that for an asymptotic observer, $\omega$ represents the observed radiation frequency \cite{Perlick:2015vta,Kobialko:2025sls}.

It is well known that when a black hole is illuminated by external sources, the boundary of its gravitational shadow is defined by the locus of spherical photon orbits—orbits with a constant radial coordinate $r = \text{const}$ from which light can escape to infinity \cite{Perlick:2021aok,Cunha:2018acu,Grenzebach:2014fha,Grenzebach:2015oea,Kobialko:2023qzo}. For such spherical orbits, or more generally for photon spheres \cite{Virbhadra:1999nm,Claudel:2000yi,Virbhadra:2002ju,Virbhadra:2022ybp}, the two conditions $\mathcal{R} = \mathcal{R}' = 0$ must be satisfied, which are equivalent to
\begin{align} \label{eq:main_v1}
A_3\omega^2-2A_4\omega L+A_5L^2+ f_r+\mathcal{C}=0, \quad
A'_3\omega^2-2A'_4\omega L+A'_5L^2+ f'_r=0.
\end{align}
Simultaneously, the condition $\mathcal{J} \geq 0$ defines the latitudinal boundaries of individual photon surfaces that host spherical orbits. The union of all such surfaces constitutes the photon region \cite{Grenzebach:2014fha,Grenzebach:2015oea}. 

Now consider a static asymptotic observer at $r=\bar{r}$, $\theta=\bar{\theta}$ with 4-velocity $v^\alpha=(0,0,N^{-1},0)$, where $N$ normalization constant determined from the condition $v^\alpha v_\alpha=-1$. We assume the spacetime to be asymptotically flat or more precisely, it has Kerr's asymptotics:
\begin{align} \label{eq:Kerr_asymptotics}
\begin{aligned}
A_1 &= r^2 + \ldots,       &\quad B_1 &= a^2 \cos^2\bar{\theta}, \\
A_2 &= r^2 + \ldots,       &\quad B_2 &= 1, \\
A_3 &= -\, r^2 + \ldots,   &\quad B_3 &= a^2 \sin^2\bar{\theta}, \\
A_4 &= -a + \ldots, &\quad B_4 &= a, \\
A_5 &= -\frac{a^2}{r^2} + \ldots, &\quad B_5 &= \frac{1}{\sin^2\bar{\theta}},
\end{aligned}
\end{align}
where $a$ Kerr spin parameter and the ellipsis denotes asymptotic higher-order contributions ($r,1/r,...$). It is important that the perturbed metrics also must not violate this asymptotic behavior. 
All specific metric examples considered in this work satisfy this condition.

The tangent vector to an observed null geodesic can be expressed as
\begin{align} \label{eq:tetrad_expand}
\dot{x}^\alpha = H\cdot e^\alpha_t + H \left(\cos \Theta e_r{}^\alpha+\sin \Theta \sin \Phi e_\theta{}^\alpha +\sin \Theta \cos \Phi e_\phi{}^\alpha\right), 
\end{align}
where $e^\alpha_t = v^\alpha$, and $e^\alpha_r$, $e^\alpha_\theta$, and $e^\alpha_\phi$ form an orthonormal tetrad \cite{Kobialko:2023qzo}, $H$ is determined from the condition $p_t=-\omega$, where photon 4-momenta reads as $p_\alpha=g_{\alpha\beta}\dot{x}^\beta$. Here, $\Phi$ and $\Theta$ represent the azimuthal and polar angles on the celestial sphere, respectively, corresponding to the direction from which the radiation arrives. We now introduce the Cartesian coordinates for the stereographic projection of the celestial sphere \cite{Kobialko:2023qzo,Bogush:2022hop}:
\begin{align} 
X = 2 \bar{r} \tan(\Theta /2) \cos \Phi,\quad Y = 2 \bar{r} \tan(\Theta /2) \sin \Phi. 
\end{align}
Here, the right-hand side has been scaled by the distance to the black hole, $\bar{r}$, to avoid dealing with numerically small values.

For an asymptotic observer, an approximation for the shadow radius $R = \sqrt{X^2 + Y^2}$ can also be derived using the approximate relation
\begin{align} 
\Theta=2 \tan(\Theta /2) =\bar{r}^{-1} R.
\end{align}
In particular, evaluating Eq. (\ref{eq:Hamilton-Jacobi}) at the observer's position and using the Kerr asymptotic (\ref{eq:Kerr_asymptotics}), and Eq. (\ref{eq:tetrad_expand}) together with the assumption of a vanishing plasma frequency at infinity ($\bar{\omega}_p = 0$), we obtain the following expression for the Carter constant and angular momentum:
\begin{align}
\mathcal{C}=\omega^2 R^2+ 2 a \omega^2 R \cos \Phi \sin \bar{\theta} +a^2\omega^2 \sin^2 \bar{\theta}, \quad  L=- \omega R \sin \bar{\theta}  \cos \Phi,
\end{align}
where $a$ Kerr spin parameter arising in the asymptotic limit. It should be noted that this expression, depends on the observer's frame choice and, for instance, differs from that defined for a Zero Angular Momentum Observer (ZAMO). However, as shown in \cite{Kobialko:2022ozq,Kobialko:2023qzo}, this difference manifests only as a shift in the chosen location of the shadow's center, leaving its intrinsic shape invariant.

Substituting these relations into Eq. (\ref{eq:main_v1}) yields the final expression for the shadow boundary:
\begin{align} \label{eq:main_v2}
&A_3+2A_4\rho R+A_5 \rho^2R^2+\omega^{-2}f_r+ R^2+ 2 a R \rho+a^2\sin^2 \bar{\theta}=0, \\
&A'_3+2A'_4\rho R+A'_5\rho^2R^2 + \omega^{-2} f'_r=0, \label{eq:main_v2_b}
\end{align}
where
\begin{align} 
\rho=  \sin \bar{\theta}  \cos \Phi.
\end{align}
Note that this implicit form is particularly convenient for constructing the perturbative expansion. Its internal consistency is verified by substituting the explicit expressions for $R$ and $\rho$ from Refs.~\cite{Kobialko:2022ozq,Kobialko:2023qzo} into Eqs.~(\ref{eq:main_v2}), (\ref{eq:main_v2_b}), which yields an identity satisfied for arbitrary metric functions $A$ with correct asymptotic behavior.

The most immediate observation is that the plasma term enters the formulas in the same manner as the metric component $A_3$. Consequently, we combine these terms by defining a unified component as follows:
\begin{align} 
A_3+\omega^{-2}f_r\rightarrow A_3.
\end{align}
Therefore, without loss of generality, we analyze the system described by the following equations:
\begin{align} 
&A_3+2A_4\rho R+A_5 \rho^2R^2+ R^2+ 2 a R \rho+a^2\sin^2 \bar{\theta}=0, \label{eq:main} \\
&A'_3+2A'_4\rho R+A'_5\rho^2R^2 =0.  \label{eq:main_a}
\end{align}
A second key observation is that the function $\rho$ remains invariant under reflection of the shadow about the image's $X$-axis, i.e., under the transformation $\sin \Phi \rightarrow -\sin \Phi$. This invariance directly implies that the shadow itself possesses reflection symmetry with respect to this axis.

In the system of equations (\ref{eq:main}) and (\ref{eq:main_a}), the unknowns are the orbital radius $r$ and the shadow radius $R$. Typically, it is more convenient to express both $R$ and the parameter $\rho$ as functions of $r$, thereby defining the shadow boundary in a parametric form. A significant drawback of this parameterization for calculating key shadow characteristics is that many of these quantities become difficult to compute in a fully analytical manner. To circumvent this issue, the present work employs perturbation theory to derive explicit expressions for $R$ directly as functions of $\rho$ or the celestial angle $\Phi$.

If an explicit expression for $R(\Phi)$ is available, then the shadow boundary in the Cartesian stereographic coordinates $(X, Y)$ is given parametrically by
\begin{align} 
X=R(\Phi) \cos\Phi, \quad Y=R(\Phi) \sin\Phi,
\end{align}
The primary characteristics of the shadow are its horizontal and vertical diameters (or sizes) \cite{Johannsen:2013vgc,Cunha:2015yba}.
\begin{align} 
D_X=X|_{\max}-X|_{min}, \quad D_Y=Y|_{\max}-Y|_{min}.
\end{align}
The second key characteristic is the position of the shadow's geometric center. In principle, several methods exist to define this center. One common approach is to compute the centroid (center of mass) of the shadow area via integration \cite{Banerjee:2019nnj,Bogush:2022hop}.
\begin{equation}
    X_{C_A} = \frac{ \int X dA }{\int dA} =
    \frac{
        2\oint R^3 \cos \Phi d\Phi
    }{
        3\oint R^2 d\Phi
    },\qquad
    Y_{C_A} = \frac{ \int Y dA }{\int dA} =
    \frac{
        2\oint R^3 \sin \Phi  d\Phi
    }{
        3\oint R^2 d\Phi
    },
\end{equation}
where $dA$ is the infinitesimal area element of the shadow. A second method is to define the center as the arithmetic mean of the boundary coordinates
\begin{align} 
X_{C_m}=\frac{1}{2}\left(X|_{\max}+X|_{min}\right), \quad
Y_{C_m}=\frac{1}{2}\left(Y|_{\max}+Y|_{min}\right).
\end{align}
In general, these two definitions yield different positions for the center. The discrepancy is most pronounced for highly deformed shadows, particularly those with significant convexity. For all subsequent quantities that depend on the choice of the center, we will use the subscript $A$ to denote the area center and $m$ for the arithmetic mean center. Note that due to the aforementioned reflection symmetry of the shadow, any consistent definition will place the center on the $X$-axis, i.e., its $Y$-coordinate will be zero.
\begin{align} 
Y_{C_{A,m}}=0.
\end{align}
Once the geometric center is determined, we can introduce a shifted polar coordinate system centered on this point.
\begin{align} 
X_{A,m}=X-X_{C_{A,m}}, \quad Y_{A,m}=Y-Y_{C_{A,m}},
\end{align}
with the corresponding new radial and angular coordinates defined as
\begin{align} 
R_{A,m}=\sqrt{X_{A,m}^2+Y_{A,m}^2}, \quad \cos \Phi_{A,m}=\frac{X_{A,m}}{R_{A,m}}.
\end{align}
In this new coordinate system, we can calculate the average shadow radius $\bar{R}_{A,m}$ as \cite{Johannsen:2013vgc,Cunha:2015yba}
\begin{align} 
\bar{R}_{A,m}=\frac{1}{2\pi}\int R_{A,m}\cdot d\Phi_{A,m}, 
\end{align}
Note that the literature also employs an alternative, area-based definition for the average shadow radius \cite{Banerjee:2019nnj,Bambi:2019tjh}
\begin{align}  \label{eq:radius_area_based}
\bar{R}=\sqrt{\frac{1}{\pi}\int dA}=\sqrt{\frac{1}{2\pi}\int R^2\cdot d\Phi}, 
\end{align}
which offers the practical advantage of being calculable directly without shifting to a coordinate system centered on the shadow.

The deviation from perfect sphericity reads as \cite{Johannsen:2013vgc,Cunha:2015yba,Banerjee:2019nnj,Bambi:2019tjh}
\begin{align} 
\delta C_{A,m}=\sqrt{\frac{1}{2\pi}\int \left(\frac{R_{A,m}-\bar{R}_{A,m}}{\bar{R}_{A,m}}\right)^2\cdot d\Phi_{A,m}}, 
\end{align}
and the mean deviation from the canonical Kerr shadow as 
\begin{align} 
\delta K_{A,m}=\sqrt{\frac{1}{2\pi}\int \left(\frac{R_{A,m}-R^{Kerr}_{A,m}}{R^{Kerr}_{A,m}}\right)^2\cdot d\Phi_{A,m}}. 
\end{align}
For these integrals, all integrand functions must be expressed in terms of the corresponding angular coordinate $\Phi_{A,m}$, or a suitable change of the integration variable must be performed. The set of characteristics $(D_X, D_Y, X_{C_{A,m}}, \bar{R}_{A,m}, \delta C_{A,m}, \delta K_{A,m})$ is widely used to provide an invariant description of a gravitational shadow \cite{Johannsen:2013vgc,Cunha:2015yba}. In the following section, we present fully analytical results for calculating these parameters using perturbation theory.

\section{Perturbation theory} 
\label{sec:perturbation}

In this work, we employ the Schwarzschild metric as the unperturbed background and introduce perturbations \cite{Vertogradov:2024dpa,Vertogradov:2024eim,Pantig:2025deu,Kobialko:2024zhc} corresponding to Kerr rotation, characterized by the spin parameter $a$, and a secondary deformation to the Kerr geometry itself, controlled by a dimensionless parameter $\epsilon$. We assume these perturbations preserve the underlying Benenti-Francavilla-type metric structure \cite{Papadopoulos:2018nvd,Benenti:1979erw}. The rotation parameter $a$ is treated as the dominant effect relative to the deformation and plasma contributions, adhering to a specific scale hierarchy $\epsilon \lesssim a^2 / M^2$. Consequently, we retain terms up to $\mathcal{O}(a^5)$ in our expansions, which ensures a cumulative error on the order of $\sim a^6$ for the calculated shadow parameters. More precisely, in our perturbative expansion we systematically retain terms 
\begin{align} 
\sim &\text{ const}, \quad \sim a, \quad \sim a^2,  \quad \sim a^3 \quad \sim a^4, \quad \sim a^5, \nonumber \\
\sim &\epsilon, \quad \sim a \epsilon, \quad \sim a^2 \epsilon, \quad \sim a^3 \epsilon, \quad \sim \epsilon^2, \quad \sim a\epsilon^2.
\end{align}
Note that including higher-order terms $\sim a^6$ and $\sim \epsilon^3$ would significantly complicate the final analytical expressions, whereas truncating the expansion at a lower order would omit information about characteristic asymmetric deformations of the shadow relative to the $Y$-axis. Nevertheless, higher-order terms may become valuable for the numerical problem of reconstructing metric parameters via optimization algorithms \cite{Kingma:2014vow}.

Accounting for the metric deformations and the Kerr-like rotation, the relevant quantities can be expanded as
\begin{align} \label{eq:decomposition}
A_3&=-\frac{\left(a^2+r^2\right)^2}{a^2-2 M r+r^2}+ M^2\sum^{3}_{i=2}\sum^{8-2i}_{j=1} A_3^{(i,j)}(r/M) \cdot\epsilon^{i-1}\cdot(a/M)^{j-1} + \mathcal{O}(a^{6}),\\
A_4&=-\frac{a \left(a^2+r^2\right)}{a^2-2 M r+r^2}+ M\sum^3_{i=2}\sum^{8-2i}_{j=2} A_4^{(i,j)}(r/M) \cdot\epsilon^{i-1}\cdot(a/M)^{j-1} + \mathcal{O}(a^{6}), \\ 
A_5&=-\frac{a^2}{a^2-2 M r+r^2}+\sum^3_{i=2}\sum^{8-2i }_{j=3} A_5^{(i,j)}(r/M) \cdot\epsilon^{i-1}\cdot(a/M)^{j-1} + \mathcal{O}(a^{6}),
\end{align}
where the leading terms must also be expanded consistently to the same order. In the sums, the upper summation limit implies the condition $(j-1) + 2(i-1) \leq 5$, which, given the scale hierarchy $\epsilon \sim a^2$, ensures a cumulative error of $\mathcal{O}(a^6)$. To prevent the deformation from introducing strong non-Kerr behavior at leading orders in the $a$-expansion, we explicitly set the coefficients $A_4^{(2,1)} = A_5^{(2,1)} = A_5^{(2,2)} = 0$. This ensures that the dominant rotational behavior in the metric originates from the Kerr contribution rather than the deformation. As noted earlier, the perturbations must also preserve the asymptotic Kerr behavior specified in Eq.~(\ref{eq:Kerr_asymptotics}). This condition is satisfied by all Kerr-like solutions considered in this work.  

The expansions for the spherical photon orbit radius $r$ and the shadow radius $R$ take the following form:
\begin{align} \label{eq:decomposition_r}
r=M\sum^3_{i=1}\sum^{8-2i}_{j=1}  r^{(i,j)} \cdot\epsilon^{i-1}\cdot(a/M)^{j-1} , \quad
R=M \sum^3_{i=1} \sum^{8-2i}_{j=1}R^{(i,j)} \cdot\epsilon^{i-1}\cdot(a/M)^{j-1}. 
\end{align}
The next step involves substituting these expansions into Eq. (\ref{eq:main}) and (\ref{eq:main_a}), expanding the resulting expressions as a series up to $\mathcal{O}(a^5)$, and solving the resulting system of linear equations for the unknown coefficient functions $r^{(i,j)}$ and $R^{(i,j)}$. To manage this algebraically intensive process, we have developed an automated computational algorithm. The resulting general expressions for $R(\Phi)$ are, however, typically cumbersome and not particularly illuminating for direct physical analysis. Therefore, we proceed to calculate directly the key invariant parameters of the shadow \cite{Johannsen:2013vgc,Cunha:2015yba}: $(D_X, D_Y, X_{C_{A,m}}, \bar{R}_{A,m}, \delta C_{A,m}, \delta K_{A,m})$.

The principal shadow parameters are fully determined by specific combinations of the metric expansion coefficients evaluated at the photon sphere of the unperturbed Schwarzschild metric, $r = 3M$ (we use the shorthand notation $A_3^{(i,j)} \equiv A_3^{(i,j)}(3)$, etc.). Specifically, the parameters governing the shadow's shape are given by the following expressions ($i=1,2$):
\begin{align}  \label{eq:S_definition}
\epsilon^{-1}\cdot S^{(i)}&\equiv243 A_5^{(2,i+2)}+108 \left(A_4^{(2,i+1)}\right)'-2 \left(A_3^{(2,i)}\right)'+6\left(A_3^{(2,i)}\right)'', \\
\epsilon^{-1}\cdot S^{\Delta}&\equiv 2/3A_3^{(2,1)}+ \left(A_3^{(2,1)}\right)' 
- 6\left(A_3^{(2,1)}\right)'' 
+ 6\left(A_3^{(2,1)}\right)''' \nonumber \\
& \quad - 54\left(A_4^{(2,2)}\right)' 
+ 162\left(A_4^{(2,2)}\right)'' 
+ 729\left(A_5^{(2,3)}\right)',
\end{align}
The parameters describing the linear dimensions of the shadow are
\begin{align} 
\epsilon^{-1}\cdot L^{(i)}_0 &\equiv A_3^{(2,i)},\\
\epsilon^{-1}\cdot L^{(i)}_1&\equiv 16 \left( \left(A_3^{(2,i)}\right)' + 9 A_4^{(2,i+1)} \right) - 3 A_3^{(2,i)}, \\
\epsilon^{-1}\cdot L^{(i)}_2&\equiv-8 \left(A_3^{(2,i)}\right)' + A_3^{(2,i)} + 18 A_3^{(2,i+2)}, 
\end{align}
Non-linear size correction reads as
\begin{align} 
\epsilon^{-2}\cdot N^{(1)}&\equiv3 \left( \left( \left( A_3^{(2,1)} \right)' \right)^2 + 36 A_3^{(3,1)} \right) + \left( A_3^{(2,1)} \right)^2, \\
\epsilon^{-2}\cdot N^{(2)}&\equiv 6 \left( A_3^{(2,1)} \right)' \left( A_3^{(2,2)} \right)' + 2A_3^{(2,1)} A_3^{(2,2)} + 108 A_3^{(3,2)},
\end{align}
The shifts of the shadow's center relative to the origin of the celestial coordinates are described by the following expressions:
\begin{align}
&\epsilon^{-1}\cdot W_1 =7 \left(A_3^{(2,1)}\right)' - 12 \left(A_3^{(2,1)}\right)'' - 108 \left(A_4^{(2,2)}\right)' + 27 \left(A_3^{(2,3)}\right)' + 243 A_4^{(2,4)}, \\
&\epsilon^{-1}\cdot W_2 =-25 \left(A_3^{(2,1)}\right)' + 30 \left(A_3^{(2,1)}\right)'' + 4 A_3^{(2,1)} + 18 \left(A_3^{(2,1)}\right)''' \nonumber\\
&+ 702 \left(A_4^{(2,2)}\right)' + 486 \left(A_4^{(2,2)}\right)'' - 108 A_4^{(2,2)} + 2187 \left(A_5^{(2,3)}\right)' + 1944 A_5^{(2,3)}, \\
&\epsilon^{-2}\cdot W_3= 3 \left(A_3^{(2,1)}\right)' \left( \left(A_3^{(2,1)}\right)'' + 9 \left(A_4^{(2,2)}\right)' \right)  + \left( \left(A_3^{(2,1)}\right)' \right)^2 + 54 \left(A_3^{(3,1)}\right)' + 486 A_4^{(3,2)}.
\end{align}
The following small parameters, constructed from the quantities defined above, also play a crucial role:
\begin{align} 
S &\equiv S^{(1)}+a S^{(2)}/M, \quad L_{0,1,2} \equiv L^{(1)}_{0,1,2}+a L^{(2)}_{0,1,2}/M, \quad N \equiv N^{(1)}+a N^{(2)}/M.
\end{align}

First, we determine the maximum and minimum values of the Cartesian coordinates $X$ and $Y$ that define the shadow boundary. This is achieved by applying the standard extremum condition.
\begin{align} 
\frac{dX}{d\Phi}\Big|_{\Phi_{X}}=0, \quad \frac{dY}{d\Phi}\Big|_{\Phi_{Y}}=0.
\end{align}
We again employ a series expansion to solve for the corresponding extremal angles.
\begin{align} \label{eq:decomposition_Phi}
\Phi_{X,Y}=\sum^{3}_{i=1}\sum^{8-2i}_{j=1}  \Phi_{X,Y}^{(i,j)} \cdot\epsilon^{i-1}\cdot(a/M)^{j-1}.  
\end{align}
Our calculations reveal that, to the considered perturbative order, the maximum and minimum of the $X$-coordinate on the shadow boundary lie on the symmetry axis at $\Phi_X = 0$ and $\Phi_X = \pi$, respectively. In contrast, the angles $\Phi_Y$ corresponding to the vertical extrema admit a non-trivial series expansion. The resulting horizontal diameter $D_X$ and vertical diameter $D_Y$ are then given by
\begin{align}
\begin{aligned}
D_X&=M\left[ 6\sqrt{3} 
-\frac{1}{\sqrt{3} }\cdot\frac{a^2}{M^2} 
- \frac{-256\sin^4\bar{\theta}+ 512\sin^2\bar{\theta}+ 304}{1728\sqrt{3} }\cdot\frac{a^4 }{M^4} \right. \nonumber \\
&\quad \left. - \frac{L_0 }{3\sqrt{3}} 
- \frac{N}{324\sqrt{3}}- \frac{ 1}{54\sqrt{3} } \left( \sin^2\bar{\theta}\left( 2S - \frac{L_0}{2} + \frac{L_1}{2} \right) + L_2 \right)\cdot\frac{a^2}{M^2}\right]+ \mathcal{O}(a^{6}), \\
D_Y&=M\left[ 6\sqrt{3} 
+ \frac{\sin^2\bar{\theta}- 1}{\sqrt{3} }\cdot\frac{a^2}{M^2} 
- \frac{ 304\sin^4\bar{\theta}- 608\sin^2\bar{\theta}+ 304}{1728\sqrt{3} }\cdot\frac{a^4 }{M^4} \right. \nonumber\\
&\quad \left. - \frac{L_0}{3\sqrt{3}} 
- \frac{N }{324\sqrt{3}}- \frac{ 1}{54\sqrt{3} } \left( \sin^2\bar{\theta}\left(\frac{L_0}{2}  + \frac{L_1 }{2}\right) + L_2 \right)\cdot\frac{a^2}{M^2}  \right]+ \mathcal{O}(a^{6}).
\end{aligned}
\end{align}
These expressions allow for a straightforward calculation of a simple measure of the shadow's asymmetry, namely the axis ratio $D_Y / D_X$. The resulting formula is remarkably compact.
\begin{align} 
D_{Y}/D_{X}=1+\frac{a^2 \sin^2 \bar{\theta}}{18 M^2}\cdot\left(1+\frac{S }{27}- \frac{35 \sin^2 \bar{\theta}-76 }{108  }\cdot \frac{a^2}{M^2} \right) +\mathcal{O}(a^6),
\end{align}
This constitutes one of the principal results of this work: the axis ratio $D_Y/D_X$ depends only on the single shape parameter $S$, defined in Eq. (\ref{eq:S_definition}).

The next key characteristic is the position of the shadow's geometric center. We adopt the area center $X_{C_A}$ for our primary analysis, as it yields a more accurate subsequent approximation for the shadow shape. For the horizontal coordinate of the shadow's center, we obtain the following expressions:
\begin{align} 
\begin{aligned}
X_{C_A}&=\frac{a\sin\bar{\theta}}{972} \left[ 1944 + \left( 10 \sin^4\bar{\theta} - 56 \sin^2\bar{\theta} + 88 \right)\cdot \frac{a^4}{M^4} + \left(1 - \frac{1}{2}\sin^2\bar{\theta}  \right) \cdot \frac{216a^2}{M^2} \right. \\
&\left. - \frac{a^2}{M^2}  \left( 4W_1 + W_2 \sin^2\bar{\theta} \right) \right. \left. -  \frac{27 }{4}\cdot ( L_1+3 L_0) -2 W_3 \right].
\end{aligned}
\end{align}
It is noteworthy that the previously defined parameters $W$ appear exclusively in the expressions for the center shifts and do not contribute to any other shadow characteristics calculated in this work. Since the position of the shadow center is highly sensitive to the choice of the observer's reference frame (tetrad) relative to the black hole's coordinate system, this characteristic is not directly observable in a model-independent way. Therefore, we will omit its explicit calculation in the specific examples that follow.

Next, we introduce a shifted polar coordinate system centered on the point $(X_{C_A}, 0)$ i.e.,
\begin{align} 
X_{A}=X-X_{C_{A}}, \quad Y_{A}=Y,
\end{align}
and
\begin{align}  \label{eq:Phi}
R_{A}=\sqrt{X_{A}^2+Y_{A}^2}, \quad \cos \Phi_{A}=\frac{X_{A}}{R_{A}}.
\end{align}
Our task now is to express the original angular coordinate $\Phi$ in terms of the new one, $\Phi_A$. This transformation is also derived using perturbation expansion. In practice, it is more straightforward to work with the expansion for $\cos\Phi$:
\begin{align} 
\cos\Phi=\sum^3_{i=1}\sum^{8-2i}_{j=1} \Phi_{A}^{(i,j)}\left(\cos \Phi_{A}\right) \cdot\epsilon^{i-1}\cdot(a/M)^{j-1},  
\end{align}
This expansion is substituted into Eq. (\ref{eq:Phi}) to determine all coefficients. Using this result, we obtain the shadow boundary $R_A(\Phi_A)$ in the coordinate system centered on $X_{C_A}$. The explicit expression is provided in Appendix, Eq. (\ref{eq:R_A}). Integrating $R_A(\Phi_A)$ over a full period then yields the average shadow radius
\begin{align} 
\begin{aligned}
\bar{R}_{A}&=M\left[ 3\sqrt{3} 
+ \frac{\sin^2\bar{\theta}- 2}{4\sqrt{3} }\cdot\frac{a^2}{M^2} 
- \frac{ -11\sin^4\bar{\theta}-24\sin^2\bar{\theta}+ 152}{1728\sqrt{3} }\cdot\frac{a^4 }{M^4} \right. \nonumber \\
&\quad \left. - \frac{L_0 }{6\sqrt{3}} 
- \frac{N}{648\sqrt{3}}- \frac{ 1}{108\sqrt{3} } \left( \sin^2\bar{\theta}\left( S + \frac{L_1}{2} \right) + L_2 \right)\cdot\frac{a^2}{M^2} \right]+ \mathcal{O}(a^{6}). 
\end{aligned}
\end{align}
For the area-based definition (\ref{eq:radius_area_based}), we find only a slight adjustment independent of the Kerr metric deformation
\begin{align} 
\bar{R}-\bar{R}_{A}=\frac{a^4 \sin ^4\bar{\theta}}{576 \sqrt{3} M^3}+\mathcal{O}(a^6).
\end{align}
As anticipated, this value is slightly overestimated, owing to the greater weight of shadow boundary regions located farther from the center.

Using the calculated average radius $\bar{R}_A$, the deviation from perfect sphericity $\delta C_A$ is readily obtained. Notably, within the perturbative framework, this characteristic also admits a compact analytical form.
\begin{align} 
\delta C_A=\frac{a^2\sin^2 \bar{\theta}}{36\sqrt{2} M^2}\left(1+\frac{S}{27}- \frac{\frac{310}{9} \sin^2 \bar{\theta}-76 }{108}\cdot \frac{a^2}{M^2} \right)+\mathcal{O}(a^6).
\end{align}
Note that to the order considered, the use of the average radius $\bar{R}$ instead of $\bar{R}_A$ leads to the same result. At the same time, using a different center $X_{C_m}$ leads to an expected increase in this deviation.
\begin{align} 
\delta C_m-\delta C_A=\frac{\sqrt{2} a^4 \sin^4\bar{\theta}}{2187 M^4}+\mathcal{O}(a^6).
\end{align}
We now introduce another significant parameter: the normalized shift between the two center definitions
\begin{align} 
\Delta X_{C}\equiv \frac{X_{C_m}-X_{C_A}}{\bar{R}_{A,m}}.
\end{align}
Our calculations yield the following results:
\begin{align}
\Delta X_{C}=\frac{a^3\sin^3 \bar{\theta}}{81\sqrt{3}M^3}\left(1- \frac{S^{\Delta}}{36}- \frac{57 \sin^2 \bar{\theta}-126 }{108}\cdot \frac{a^2}{ M^2}\right)+\mathcal{O}(a^6),
\end{align}
and
\begin{align}
\begin{aligned}
\frac{R_A}{\bar{R}_A}  =&1 - \frac{a^2 \sin^2\bar{\theta}}{36M^2} 
\left(1 - \frac{19\sin^2\bar{\theta} - 38}{54 }\cdot \frac{a^2}{M^2} + \frac{S}{27} \right)\cdot  \cos 2\Phi_A \\
& + \frac{ a^3\sin^3\bar{\theta}}{81 \sqrt{3} M^3} 
\left(1 - \frac{63 \sin^2\bar{\theta} - 126}{108}\cdot \frac{a^2}{M^2} - \frac{S^{\Delta} }{36}\right)\cdot \cos 3\Phi_A \\
& 
- \frac{23 a^4  \sin^4\bar{\theta} }{9 \cdot 1728 M^4}\cdot \cos 4\Phi_A  + \frac{ a^5  \sin^5\bar{\theta} }{4 \cdot 729 \sqrt{3} M^5} \cdot (\cos\Phi_A + \cos 5\Phi_A)+\mathcal{O}(a^6).
\end{aligned}
\end{align}
The last expression effectively decomposes the first Fourier modes of the shadow's relative deviation from a perfect circle. It reveals, for instance, that the parameter $S^\Delta$ governs specific asymmetric deformations, while the parameter $S$ controls a symmetric stretching or compression. This distinction becomes most evident when comparing the resulting shadow shape with that of the canonical Kerr solution.
\begin{align}
\frac{R_A}{\bar{R}_A}- \frac{R^{\text{Kerr}}_A}{\bar{R}^{\text{Kerr}}_A}  =& - \frac{a^2 \sin^2\bar{\theta}}{27 \cdot 36 \cdot M^2}\left(
S \cdot  \cos 2\Phi_A  + \frac{ a\sin\bar{\theta}}{3 \sqrt{3} M}\cdot S^{\Delta}\cdot \cos 3\Phi_A\right) +\mathcal{O}(a^6).
\end{align}
Consequently, the parameter $S$ governs the symmetric deformation at the polar angles $\Phi_A = \pi/2$ and $3\pi/2$, while $S^{\Delta}$ controls the antisymmetric deformation at the polar angles $\Phi_A = \pi/4$ and $3\pi/4$.
\begin{align}
\frac{R_A}{\bar{R}_A}- \frac{R^{\text{Kerr}}_A}{\bar{R}^{\text{Kerr}}_A} \Big|_{\Phi_A=\pi/2}=\frac{a^2 \sin^2\bar{\theta} \cdot S}{27 \cdot 36 \cdot M^2}, \quad \frac{R_A}{\bar{R}_A}- \frac{R^{\text{Kerr}}_A}{\bar{R}^{\text{Kerr}}_A} \Big|_{\Phi_A=\pi/4, 3\pi/4}=\pm\frac{a^3 \sin^3\bar{\theta} \cdot S^{\Delta}}{27 \cdot 36 \cdot 3 \sqrt{6} \cdot M^3}.
\end{align}

Finally, we calculate the mean deviation from the canonical Kerr shadow, denoted $\delta K$, ensuring both shadows are evaluated relative to the same geometric center.
\begin{align}
\begin{aligned}
\delta K_{A}&=\pm\frac{1}{1944} \left[36 L_0 + 2\left(L_0 + L_2\right)\cdot \frac{a^2}{M^2}+ \left(2S+L_1 -L_0\right)\cdot \frac{a^2 \sin^2 \bar{\theta}}{M^2} + \frac{N}{3} \right] +\mathcal{O}(a^6),
\end{aligned}
\end{align}
where $+$ for $A_3^{(2,1)}>0$. 

Thus, all principal characteristics of the gravitational shadow \cite{Johannsen:2013vgc,Cunha:2015yba} have been derived as relatively simple polynomial expressions. This formulation enables a fully analytical investigation of how various metric parameters influence the shadow's observable features and help with problem of recovering these parameters from observational data using standard optimization techniques what will be discussed in more detail below.

The perturbative framework may also be worth exploring for an attempt to calculate other key lensing observables, such as image magnifications and distortion parameters \cite{Virbhadra:2022iiy,Virbhadra:2024xpk,Virbhadra:2007kw}. These quantities are central to Virbhadra's conjecture on the vanishing sum of image distortions \cite{Virbhadra:2022iiy}. Testing this in a plasma is notoriously difficult, as it demands extreme ray-tracing precision. Similarly, attempting to apply it to studies of image compactness \cite{Virbhadra:2022ybp} represents a challenging yet potentially fruitful direction for future work.

\section{Examples} 
\label{sec:examples} 

The following examples serve to demonstrate the effectiveness and accuracy of the perturbative expansions derived above. 

\subsection{Kerr-Newmann Black Hole}

The components of the Kerr-Newman metric in the Benenti-Francavilla form \cite{Papadopoulos:2018nvd} are given by
\begin{align} 
A_3=-\frac{f^2}{\Delta } \quad A_4=-\frac{a f}{\Delta }, \quad
A_5=-\frac{a^2}{\Delta }, 
\end{align}
where
\begin{align} 
f&=r^2+a^2,\quad \Delta=r^2+a^2-2 M r+Q^2.
\end{align}
We choose the dimensionless charge squared, $\epsilon = q^2 \equiv Q^2/M^2$, as the small deformation parameter. In this context, our adopted scale hierarchy $\epsilon \lesssim a^2/M^2$ translates to $Q^2 \lesssim a^2$. Consequently, our perturbative framework remains applicable even to solutions where the spin and charge parameters are of comparable magnitude, provided both are sufficiently small.

The corresponding expansion coefficients are (we include $\epsilon$ in the definition of $A^{(i,j)}$)
\begin{align}
&A_3^{(2,1)}(r) = \frac{q^2r^2}{(r-2)^2}, \quad 
A_3^{(2,3)}(r) = -\frac{4q^2}{(r-2)^3}, \quad
A_3^{(3,1)}(r) = -\frac{ rq^4}{(r-2)^3},\\
&A_4^{(2,2)}(r)= \frac{q^2}{(r-2)^2}, \quad 
A_4^{(2,4)}(r)= -\frac{(r+2)q^2}{(r-2)^3 r^2}, \quad
A_4^{(3,2)}(r)= -\frac{q^4}{(r-2)^3 r}, \\
&A_5^{(2,3)}(r)= \frac{q^2}{(r^2 - 2r)^2}.
\end{align}
while all other coefficients vanish. For the key parameters governing the shadow's shape and size, we obtain
\begin{align}
S=27q^2,\quad S^\Delta=-54q^2, \quad L_0=9q^2, \quad L_1=-75q^2, \quad L_2=33q^2, \quad N = 189q^4.
\end{align}

\begin{figure}[tb!]
\centering
  \subfloat[][]{
  \includegraphics[scale=0.5]{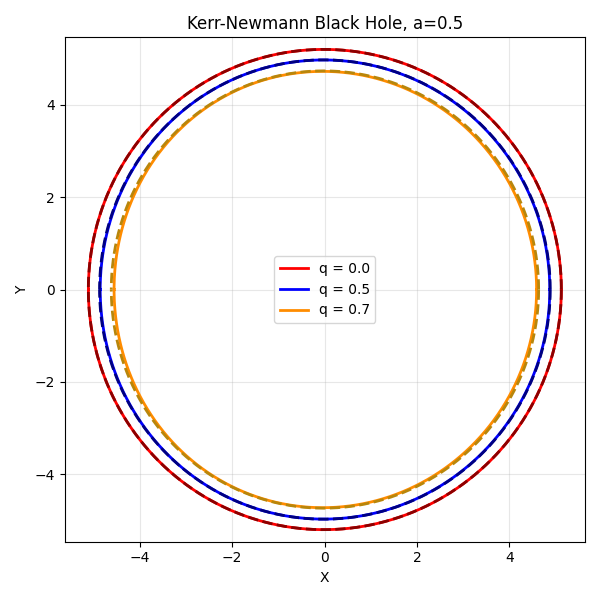} \label{shb0}
 }
  \subfloat[][]{
  \includegraphics[scale=0.5]{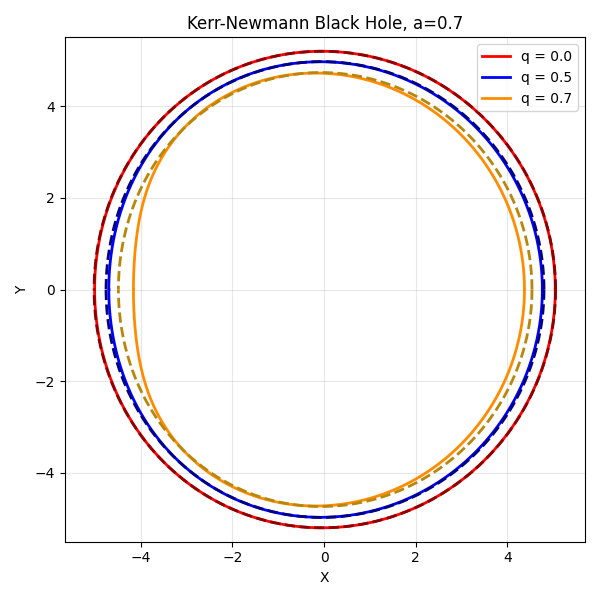} \label{shb1}
 }\\
   \subfloat[][]{
  \includegraphics[scale=0.5]{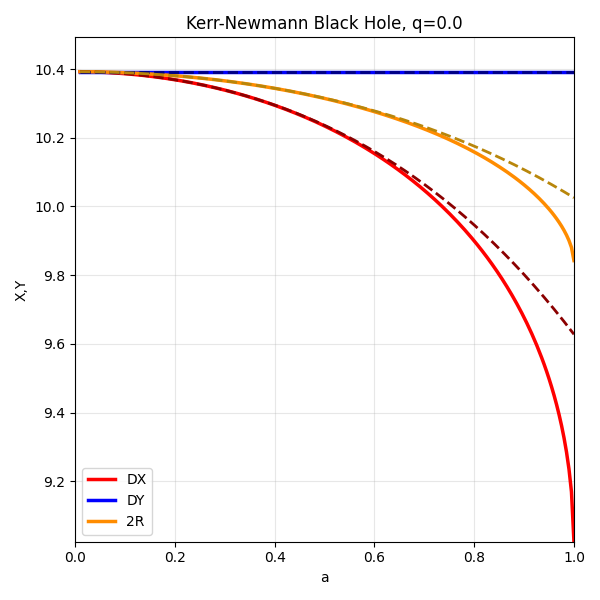} \label{shb2}
 }
    \subfloat[][]{
  \includegraphics[scale=0.5]{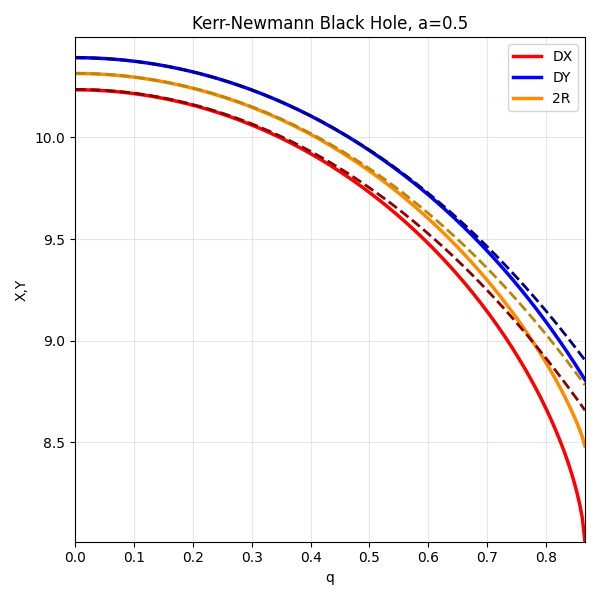} \label{shb4}
 }
\caption{Gravitational shadow for a Kerr-Newman black hole, observed at an inclination angle $\bar{\theta} = \pi/2$, for various combinations of the spin $a$ and charge $Q$ parameters. In Figs. \ref{shb0} and \ref{shb1}, the shadow shape is shown in coordinates centered on the shadow's geometric center. The exact shape is depicted by solid lines, while the fifth-order approximation is indicated by dotted lines. The dependence of the principal size parameters ($D_X$, $D_Y$, $2\bar{R}_A$ ) on the spin $a$ and the dimensionless charge $q = Q/M$ is illustrated in Figs. \ref{shb2} and \ref{shb4}, respectively. }
\label{flux_0}
\end{figure}

\begin{figure}[tb!]
\centering
  \subfloat[][]{
  \includegraphics[scale=0.5]{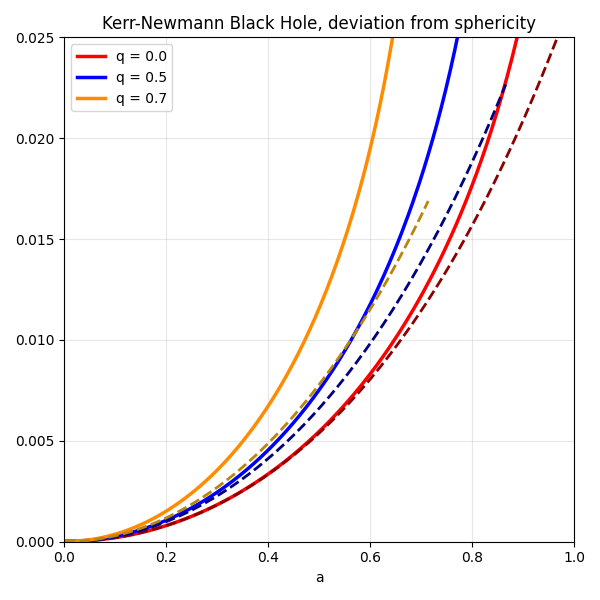} \label{plot}
 }
\caption{The deviation from perfect sphericity $\delta C_A$ for a Kerr-Newman black hole as a function of the spin parameter $a$, shown for different values of the dimensionless charge $q = Q/M$. Solid lines represent the exact values, while dotted lines correspond to the fifth-order approximation.}
\label{flux_1}
\end{figure}

To illustrate the accuracy of the derived approximation, we first recall the exact parametric expression for the Kerr-Newman shadow \cite{Kobialko:2022ozq,Kobialko:2023qzo}:
\begin{align}
X_E&=\frac{a^2 (M+r)+r \left(r (r-3 M)+2 Q^2\right)}{\sin\bar{\theta}\cdot a (r-M)}, \quad \cos \Phi_E = X_E/R_{\text{E}}, \\
R^2_{\text{E}} &=\frac{2 a^2 \left(M^2+r^2\right)+2 r \left(r^3-3 M^2 r+2 M Q^2\right)}{(r-M)^2}-a^2 \sin ^2\bar{\theta},
\end{align}
For comparison, we plot both the exact parametric curve and our fifth-order approximation.
\begin{align}
X_E,Y_E=(R_E \cos \Phi_E- C_A,R_E \sin \Phi_E), \quad X,Y=(R_A \cos \Phi_A,R_A \sin \Phi_A).
\end{align}
Figure \ref{flux_0} demonstrates that our approximation performs well across a considerable range of parameters, including cases where $Q^2 > a^2$ or where $a \sim 0.5 M$ and $Q \sim 0.5 M$. The largest deviations occur only for the extreme solution $Q^2 + a^2 = M^2$, where higher-order corrections in $\epsilon$ and $a$ become significant. The vertical diameter $D_Y$ is approximated with the highest accuracy, even for large parameter values. In contrast, shape characteristics such as the deviation from sphericity $\delta C$ diverge from the exact results more rapidly, as shown in Fig. \ref{flux_1}. 

In explicit form, deviation from sphericity and from the Kerr metric reads as
\begin{align} 
\delta C_A&=\frac{a^2\sin^2 \bar{\theta}}{36\sqrt{2}}\left(1+\frac{Q^2}{M^2}- \frac{\frac{310}{9} \sin^2 \bar{\theta}-76 }{108}\cdot \frac{a^2}{M^2} \right)+\mathcal{O}(a^6), \\
\delta K_{A}&= \frac{7 \cdot Q^2}{M^2} \left(\frac{1}{42} -\frac{5 a^2 \sin^2\bar{\theta}}{21 \cdot108 \cdot M^2} +  \frac{a^2}{162 \cdot M^2}   + \frac{Q^2}{216 \cdot M^2}\right)+\mathcal{O}(a^6).
\end{align}

The primary advantage of having explicit analytical expressions for the shadow boundary is the significant simplification they provide for the inverse problem of extracting metric parameters from an observed shadow \cite{Banerjee:2019nnj,Tsukamoto:2014tja}. In a general $r$-parametric approach, one must repeatedly map the orbital radius $r$ to the celestial angle $\Phi$ and shadow radius $R$ for each new set of trial parameters, as the mapping $r\rightarrow (\Phi, R)$ is parameter-dependent. Reconstructing parameters would then require constructing a large, pre-computed mapping or training a neural network over a broad parameter space—a computationally expensive process. In contrast, with an explicit expression $R(\Phi)$ available, one can directly formulate the fitting procedure as minimizing a loss function, e.g., $\sum \left( R_{\text{obs}}(\Phi_i) - R_{\text{model}}(\Phi_i) \right)^2$, using a standard optimizer such as Adam \cite{Kingma:2014vow}. Leveraging this, we have developed a parameter estimation algorithm that employs our fifth-order analytical approximation to reconstruct the spin $a$ and the dimensionless charge $q$ of a Kerr-Newman black hole. As synthetic observational data, we used the exact parametric expressions for the shadow boundary. Predicted metric parameters for a fixed observer inclination $\bar{\theta} = \pi/2$ are presented in Table \ref{tab:placeholder}, demonstrating satisfactory accuracy. However, for other inclination angles $\bar{\theta}$, the results are less promising, as a higher-order perturbative expansion would be required to maintain accuracy. 

\begin{table}[]
    \centering
    \begin{tabular}{|c|c|c|c|c|c|c|c|}
    \hline
      Example, $(a,q)$ & $0.2,0.0$ & $0.5,0.0$ & $0.9,0.0$ & $0.2,0.2$& $0.5,0.3$ & $0.5,0.5$ & $0.7,0.3$ \\
        \hline
      Predicted, $(a,q)$ & $0.197,0.007$ & $0.501,0.000$ & $0.970,0.000$ & $0.196,0.200$ & $0.506,0.299$ & $0.526,0.500$ & $0.731,0.296$ \\
        \hline
    \end{tabular}
    \caption{Predicted Kerr-Newman metric parameters for a fixed observer inclination $\bar{\theta} = \pi/2$. The values were obtained by minimizing the loss function $\sum \left( R_{\text{obs}}(\Phi_i) - R_{\text{model}}(\Phi_i) \right)^2$, where $R_{\text{obs}}$ is the exact shadow radius and $R_{\text{model}}$ is the fifth-order approximation from Eq. (\ref{eq:R_A}), using the Adam optimization algorithm. }
    \label{tab:placeholder}
\end{table}

\subsection{Kerr-Newmann  Black Hole in a plasma medium}

The key requirement for shadow spectroscopy \cite{Pantig:2025deu} is a frequency-dependent shift in the observed gravitational shadow. This condition is naturally satisfied in two scenarios: observations involving beams of massive particles \cite{Kobialko:2024zhc}, or radiation propagating through a plasma medium \cite{Perlick:2015vta}. In this work, we focus on and demonstrate the applicability of our analytical results to the latter case.

The components of Kerr-Newmann metric in the Benenti-Francavilla form taking into account the power law of plasma distribution are
\begin{align} 
A_3=-\frac{f^2}{\Delta } +\omega^{-2}\beta M^\alpha\cdot  r^{-\alpha+2}, \quad A_4=-\frac{a f}{\Delta }, \quad
A_5=-\frac{a^2}{\Delta }, \quad \alpha>0.
\end{align}
where $\omega$ is the observed emission frequency, and $\alpha$ and $\beta$ are model-specific constants and
\begin{align} 
f&=r^2+a^2,\quad \Delta=r^2+a^2-2 M r+M^2q^2.
\end{align}
The metric expansion coefficients have the form
\begin{align}
&A_3^{(2,1)}(r) = \frac{q^2r^2}{(r-2)^2}+\omega^{-2}\beta\cdot  r^{-\alpha+2}, \quad 
A_3^{(2,3)}(r) = -\frac{4q^2}{(r-2)^3}, \quad
A_3^{(3,1)}(r) = -\frac{ rq^4}{(r-2)^3},\\
&A_4^{(2,2)}(r)= \frac{q^2}{(r-2)^2}, \quad 
A_4^{(2,4)}(r)= -\frac{(r+2)q^2}{(r-2)^3 r^2}, \quad
A_4^{(3,2)}(r)= -\frac{q^4}{(r-2)^3 r}, \\
&A_5^{(2,3)}(r)= \frac{q^2}{(r^2 - 2r)^2}.
\end{align}
with all other expansion coefficients vanishing. For the principal parameters governing the shadow's shape and size, we obtain
\begin{align}
S&=27q^2+ 2 \cdot 3^{-\alpha+1} (\alpha - 2) \alpha \beta\omega^{-2} ,\quad
S^\Delta=- 54q^2+3^{-\alpha} \alpha (11 - 2\alpha^2) \beta\omega^{-2},  \\  L_1 &= -75 q^2-3^{1-\alpha } (16 \alpha -23) \beta \omega ^2, \quad L_2=33q^2+3^{1-\alpha } (8 \alpha -13) \beta\omega^{-2}, \\
L_0&=9q^2+3^{2-\alpha } \beta \omega^{-2}, \quad N=27 \left( 7q^4+ 2 \cdot 3^{-\alpha} (4\alpha - 5) \beta q^2\omega^{-2} +9^{-\alpha} ((\alpha - 4)\alpha + 7) \beta^2\omega^{-4} \right). 
\end{align}
By substituting these coefficient values into the general formulas of Sec.~\ref{sec:perturbation}, any shadow characteristic of interest can be readily computed. For instance, the deviation from perfect sphericity is given by
\begin{align} 
\delta C_A=\frac{a^2\sin^2 \bar{\theta}}{36\sqrt{2}M^2}\left(1+q^2+2 \cdot 3^{-\alpha-2} (\alpha - 2) \alpha \beta\omega^{-2} +  \frac{\frac{310}{9} \sin^2 \bar{\theta}-76 }{108}\cdot \frac{a^2}{M^2} \right)+\mathcal{O}(a^6).
\end{align}
A key feature of these analytical formulas is the explicit dependence of the shadow morphology on the observational frequency $\omega$. For instance, from the behavior of $\delta C_A$ alone, one can directly estimate parameters like $\alpha$ or $\beta$ based on the slope of $\delta C_A$ plotted against $\omega^{-2}$. This frequency dependence lays the foundation for shadow spectroscopy, which allows for the efficient disentanglement and extraction of individual metric parameters from multi-frequency observations \cite{Kobialko:2024zhc}.

\subsection{Modified Kerr and Kerr-Sen black holes}

Finally, we apply our formalism to more general metric families. One such multi-parameter family encompasses both Modified Kerr and Kerr-Sen black holes \cite{Papadopoulos:2018nvd}. The corresponding metric components in the Benenti-Francavilla form are given by
\begin{align} 
A_3=-\frac{f^2}{\Delta } \quad A_3=-\frac{a f}{\Delta }, \quad
A_5=-\frac{a^2}{\Delta },
\end{align}
where
\begin{align} 
f&=r^2+2  (M b) r  +a^2,\quad \Delta=r^2+a^2-2 r (M-M b )-\frac{H M^3 }{r}+M^2 q^2 .
\end{align}
The corresponding expansion coefficients are
\begin{align}
A_3^{(2,1)}(r)& = -\frac{r \left(H-q^2r+2 b (r-4) r\right)}{(r-2)^2}, \quad 
A_3^{(2,3)}(r) = \frac{4 \left(H-r \left(4 b+q^2\right)\right)}{(r-2)^3 r}, \\
A_3^{(3,1)}(r) &= -\frac{\left(H-r \left(4 b+q^2\right)\right)^2}{(r-2)^3 r}, \quad A_5^{(2,3)}(r)= \frac{4 b r-H+Q^2 r}{(r-2)^2 r}, \quad A_4^{(2,2)}(r)= \frac{r \left(4 b+q^2\right)-H}{(r-2)^2 r},\\
A_4^{(2,4)}(r)&= \frac{H (r+2)-r \left(8 b r+q^2 (r+2)\right)}{(r-2)^3 r^3}, \quad
A_4^{(3,2)}(r)= \frac{\left(H-r \left(4 b+q^2\right)\right) \left(r \left(2 b r+q^2\right)-H\right)}{(r-2)^3 r^3}.
\end{align}
with all other coefficients being zero. For the key parameters governing the shadow's shape and size, we obtain
\begin{align}
S&=54 b-19 H+27 q^2,\quad S^\Delta=-6 \left(18 b-7 H+9 q^2\right), \\ L_0&=18 b-3 H+9 q^2, \quad L_1=-150 b+41 H-75 q^2, \quad L_2=66 b-19 H+33 q^2, \\ N &= 3 \left(144 b^2+84 b \left(3 q^2-2 H\right)+16 H^2-66 H q^2+63 q^4\right).
\end{align}
In particular for positive values of the parameter $H$, in contrast to other deformation parameters considered previously, lead to an overall increase in the shadow's linear dimensions. The deviation from perfect sphericity $\delta C$ retains a relatively simple analytical form.
\begin{align} 
\delta C_A&=\frac{a^2\sin^2 \bar{\theta}}{36\sqrt{2} M^2}\left(1+q^2+2b-\frac{19 H}{27}- \frac{\frac{310}{9} \sin^2 \bar{\theta}-76 }{108}\cdot \frac{a^2}{M^2} \right)+\mathcal{O}(a^6).
\end{align}
It should be noted that the same optimization method used successfully to reconstruct the charge in Kerr-Newman metrics proves less effective for disentangling individual parameters in more general multi-parameter families. A similar limitation has been observed in static spacetimes \cite{Kobialko:2024zhc,Pantig:2025deu}. In such cases, multi-frequency observations can significantly enhance the robustness of parameter recovery. We anticipate that the analytical formulas derived in this work will facilitate future investigations of this challenging problem.

\section{Conclusion} 

We have presented an analytical framework for calculating the key parameters of a gravitational shadow—such as its horizontal and vertical diameters ($D_X$, $D_Y$), the coordinate of its center ($X_C$), the average radius ($\bar{R}$), the deviation from perfect sphericity ($\delta C$), and the mean deviation from the Kerr shadow ($\delta K$)—using perturbation theory. This method yields these parameters as simple polynomial expressions accurate to $\mathcal{O}(a^5)$, where $a$ is the Kerr spin parameter, thereby bypassing the need for repeated numerical integration of parametric equations.

Our results demonstrate high accuracy, particularly for approximating the shadow's size, with somewhat lower but still satisfactory precision for its detailed shape. This analytical formulation serves a dual purpose: it facilitates a direct analysis of how individual metric parameters influence the shadow's observable features, and it significantly simplifies the inverse problem of reconstructing these parameters from observational shadow data. As a practical illustration, we have shown that our expressions enable efficient parameter estimation via standard optimization algorithms like Adam.

A limitation arises in multi-parameter families, where the accuracy of parameter recovery can diminish. In such scenarios, we anticipate that multi-frequency observations combined with the shadow spectroscopy technique can significantly enhance the robustness of parameter inference. A key feature of our formalism is its inherent incorporation of plasma dispersion effects, making it directly applicable to and highly relevant for the analysis of multi-frequency astrophysical observations.

\begin{acknowledgments}
The work was supported by the Foundation for the Advancement of Theoretical Physics and Mathematics ’BASIS’.
\end{acknowledgments}

\appendix

\section{Shadow radius} 
\label{eq:AppA}

Approximate expression for the gravitational shadow radius in coordinates centered on its geometric center.
\begin{align}
R_A = M &\left[ \frac{1}{\sqrt{3}} \left( 9 
- \frac{19 a^4}{216 M^4} + \frac{11 a^4 \sin^4\bar{\theta}}{1728 M^4} - \frac{a^2}{2 M^2}+ \sin^2\bar{\theta} \left( \frac{3 a^4}{216 M^4} + \frac{a^2}{4 M^2} \right) \right. \right. \nonumber\\
&\quad \left. -  \frac{ a^2 \sin^2\bar{\theta} }{216 M^2}(L_1 + 2S) - \frac{a^2 L_2 }{108 M^2}    - \frac{L_0 }{6}- \frac{N }{648}\right) \nonumber\\
&\quad + \frac{  a^2 \sin^2\bar{\theta}}{432\sqrt{3} M^2} \left( \frac{35 a^2 \sin^2\bar{\theta}}{M^2} - \frac{70 a^2}{M^2} + 2L_0- 4 S  - 108 \right)\cdot \cos 2\Phi_A\nonumber \\
&\quad + \frac{  a^3 \sin^3\bar{\theta}}{2916 M^3} \left( -\frac{60 a^2 \sin^2\bar{\theta}}{M^2} + \frac{120 a^2}{M^2} - 2L_0  - 3S^\Delta  + 108 \right)\cdot \cos 3\Phi_A \nonumber\\
&\quad \left. - \frac{23 \cdot a^4 \sin^4\bar{\theta}}{1728\sqrt{3} M^4} 
\cdot \cos 4\Phi_A+  \frac{ a^5 \sin^5\bar{\theta}}{972 M^5} \cdot (\cos\Phi_A + \cos 5\Phi_A)\right]. \label{eq:R_A}
\end{align}

\bibliography{main}

\end{document}